\documentclass[reprint,longbibliography,preprintnumbers,nofootinbib,amsmath,amssymb,aps,prd,twocolumns]{revtex4-2}
\usepackage{dblfloatfix}
\usepackage{graphicx}
\usepackage{subfigure}
\usepackage{dcolumn}
\usepackage{bm}
\usepackage{color}

\usepackage{amssymb}
\usepackage{pifont}
\definecolor{green2}{rgb}{0.0, 0.5, 0.0}
\usepackage[dvipsnames]{xcolor}
\usepackage[colorlinks = true,
            linkcolor = blue,
            urlcolor  = blue,
            citecolor = green,
            anchorcolor = blue]{hyperref}
\usepackage{verbatim}
\makeatletter
\usepackage{multirow,graphicx,amssymb,url,mathrsfs,amsmath}
\usepackage{amsxtra,amstext,latexsym,dsfont,amsfonts}
\usepackage{color,eucal}
\usepackage[dvipsnames]{xcolor}
\usepackage{float}
\usepackage{colortbl}
\usepackage{multirow}
\usepackage{tikz}
\usetikzlibrary{tikzmark}
\usetikzlibrary{arrows}

\newcommand{\Rmnum}[1]{\expandafter\@slowromancap\romannumeral #1@}
\usepackage{tabularx, array}
\newcolumntype{L}[1]{>{\raggedright\arraybackslash}p{#1}}
\newcolumntype{C}[1]{>{\centering\arraybackslash}p{#1}}
\newcolumntype{R}[1]{>{\raggedleft\arraybackslash}p{#1}}
\makeatother

\makeatletter
\newcommand{\myBig}{\bBigg@{1.75}}
\makeatother

\begin{document}
\title{Various phase transitions in a holographic p-wave superfluid model with nonlinear terms}
\author{Yue-Peng Wang}
\affiliation{Center for gravitation and astrophysics, Kunming University of Science and Technology, Kunming 650500, China}
\author{Zi-Qiang Zhao}
\affiliation{Key Laboratory of Cosmology and Astrophysics (Liaoning), College of Sciences, Northeastern University, Shenyang 110819, China}
\affiliation{Center for gravitation and astrophysics, Kunming University of Science and Technology, Kunming 650500, China}
\author{Hui Zeng}
\email{zenghui@kust.edu.cn}
\email{Corresponding author.}
\affiliation{Center for gravitation and astrophysics, Kunming University of Science and Technology, Kunming 650500, China}
\author{Zhang-Yu Nie}
\email{niezy@kust.edu.cn}
\email{Corresponding author.}
\affiliation{Center for gravitation and astrophysics, Kunming University of Science and Technology, Kunming 650500, China}
\date{\today}

\begin{abstract}
This study investigates various phase transitions, including those of 2nd, 1st, and 0th order, in a holographic p-wave superfluid model incorporating 4th- and 6th-order nonlinear terms with coefficients $\lambda$ and $\tau$. We demonstrate that these nonlinear terms provide universal control over the phase transitions of the p-wave model, qualitatively consistent with findings in the holographic s-wave case. By analyzing the condensate and free energy behavior across typical phase transitions, we quantitatively map out the $\lambda-\tau$ parameter space that characterizes different transition types. For a slightly negative $\lambda$, we further establish a $\tau-\rho$ phase diagram featuring a line of first-order phase transition points that terminates at a critical point, beyond which lies a supercritical region. Our results confirm the precise tunability of the p-wave superfluid phase transitions through $\lambda$ and $\tau$. The comprehensive phase diagrams and quantitative transition criteria we provide offer a valuable resource for future studies.
\end{abstract}
\maketitle
\tableofcontents

\section{Introduction}\label{Introduction}
The Anti-de Sitter/Conformal Field Theory (AdS/CFT) correspondence\cite{Witten:1998wy,Witten:1998qj} is a transformative framework that is expected to be a powerful tool for investigating strongly coupled systems. It is also called the holographic duality because it maps the gravitational theories in higher dimensional spacetime to the quantum field theory systems in lower dimensional spacetime. Among the various applications, the holographic superfluid model realized spontaneous symmetry breaking by the scalar hair formation in the asymptotic AdS space time which is dual to the superfluid phase transition in the field theory description~\cite{Gubser:2008px,Hartnoll:2008vx}.

Early studies on the holographic superfluids mainly focus on the condensate of the charged scalars which is dual to the s-wave superfluid orders\cite{Albash:2008eh,Nakano:2008xc,Roberts:2008ns,Herzog:2008he,Hartnoll:2008kx,Horowitz:2008bn}. However, real-world phase transition processes often involve more general orders~\cite{RevModPhys.47.331,Volhardt-Wolfle-1990,PhysRevB.37.3664,1995Muller,BealMonod1996PhysRevB}. Studies on holographic superfluid systems have also been extended to include more general order parameters such as the p-wave~\cite{Gubser:2008wv,Cai:2013aca} and d-wave ones~\cite{Chen:2010mk,Benini:2010pr}, which promotes the holographic understanding of various superfluid phase transition phenomena. 

In the study of p-wave superfluid models, Ref.~\cite{Gubser:2008wv} pioneered the coupling of a vector order to a U(1) gauge field via the SU(2) gauge fields in bulk, which represents the condensate of an anisotropy order parameter. Refs.\cite{Ammon:2009xh,Cai:2010zm} further realized the anisotropic imprints on the spacetime metric. Ref.~\cite{Cai:2013aca} implemented a more convenient holographic p-wave superfluid model using the charged complex vector fields, and investigated the phase transitions with different charges and conformal dimensions, revealing a more rich phase structure. Meanwhile, Refs.~\cite{Chen:2010mk,Benini:2010pr} conducted holographic d-wave superfluid models, further complement the theoretical framework of holographic superfluid phase transitions. 

Based on the above important progresses, the competition and coexistence between the various order parameters are studied in the holographic models\cite{Basu:2010fa,Cai:2013wma,Nie:2013sda,Nie:2014qma,Nie:2015zia,Li:2014wca,Nishida:2014lta,Wang:2016jov,Arias:2016nww,Liu:2015zca,Conti:2014uda,Momeni:2013bca,Amado:2013lia}. The complex phase diagrams and novel physical phenomena of multi-codnesate have been vividly revealed, which has injected new vitality into the development of holographic superfluid models.

Recently, the role of nonlinear potential terms in holographic superfluid phase transitions has attracted increasing attention. The early work by Herzog~\cite{Herzog:2010vz} analytically investigated the effect of a quartic potential term on the condensate curve near the critical point. Ref.~\cite{Zhang:2021vwp} considered the influence of fourth power potential terms on the phase transitions within a holographic model containing both s-wave and p-wave order parameters. Ref.~\cite{Zhao:2022jvs} introduced fourth and sixth power nonlinear terms into the s-wave superfluid model, demonstrating that these nonlinear terms enable fine-tuning on the phase transitions, from the most common second-order phase transitions to 0th and 1st order ones in the probe limit. This provides novel approaches and methodologies for future studies on superfluid phase transitions, such as the realization of non-equilibrium evolution in first-order phase transitions\cite{Zhao:2023ffs}, as well as the investigation of the universality of supercritical superfluids\cite{Zhao:2024jhs,Cao:2024irr}. Universal control of the 4th and 6th power terms is also confirmed in the holographic s-wave superfluid beyond the probe limit~\cite{Zhao:2025tqq}.  

The 4th power terms are also introduced in recent studies on the holographic s+d model~\cite{Liao:2025knf}, where the control of the 4th power terms show qualitatively the same law as in the holographic s+p model~\cite{Zhang:2021vwp}. Therefore, it is expected that the control from the 4th and 6th power terms on more general condensates follows the same universal laws as in the case of s-wave order~\cite{Zhao:2022jvs}. To confirm the above conjecture, it is necessary to test the universal control of the 4th and 6th power terms in a holographic p-wave superfluid model. In addition, taking advantage of the powerful nonlinear terms, we are able to realize various phase transitions in the p-wave model in the probe limit, where the order parameter is anisotropic. This will contribute significantly to further investigations on the dynamical stability as well as non-equilibrium evolution of anisotropic superfluid configurations. Furthermore, based on the discoveries in the holographic s-wave model with 4th and 6th power terms, the stable isotropic bubble configuration is realized in the spinodal region of a first order phase transition~\cite{Zhao:2022jvs,Zhao:2023ffs}. The extended study in the p-wave model will introduce anisotropic domain wall and bubble configurations which is important in understanding the nature of membranes and films.

In this paper, we will investigate the impact of 4th and 6th power nonlinear terms on the phase transitions of a holographic p-wave superfluid model in the probe limit. By tuning the coefficients \(\lambda\) and \(\tau\) of the nonlinear terms, we expect universal and precise control on the phase transitions between 2nd order, 1st order and 0th order ones. We will delineate the parameter space dominated by the various phase transitions for convenient reference in future studies. The remaining sections of this article are organized as follows: The setup of the holographic model including the detailed equations and formula is introduced in Section \ref{secSetup}. The condensate and free energy curves of the various phase transitions are presented in Section \ref{secPhaseTransition}, which confirm the universal control of the 4th and 6th power terms. We also plot the $\tau-\lambda$ parameter space for the various phase transitions Section \ref{secPhaseTransition}. In Section \ref{secParameterSpace}, to illustrate a useful application of the universal control from 4th and 6th power terms, we give a phase diagram including the line of first order phase transition points terminating at a critical point beyond which is the domain of the supercritical region. Finally, in Section \ref{secConclusion}, we summarize the results and discuss interesting topics for future studies. 
\section{The holographic setup}\label{secSetup}
We consider the complex vector field holographic p-wave model~\cite{Cai:2013aca,Cai:2013pda} with 4th and 6th power nonlinear self-interaction terms in the (3+1) dimensional asymptotic AdS spacetime. The bulk action of this model is given by
\begin{align}
S=&\,S_{M}+S_{G}~,\label{Lagall}\\
S_G=&\,\frac{1}{2\kappa_g ^2}\int d^{4}x\sqrt{-g}(R-2\Lambda)~,\label{Lagg}\\
S_M=&\int d^{4}x\sqrt{-g}\Big(-\frac{1}{4}F_{\mu\nu}F^{\mu\nu}
-\frac{1}{2}\rho_{\mu\nu}^{\dagger}\rho^{\mu\nu}-m_p^2\rho^\dagger_{\mu}\rho^{\mu} \nonumber\\
&-\lambda(\rho^\dagger_{\mu}\rho^{\mu})^{2}-\tau(\rho^\dagger_{\mu}\rho^{\mu         })^{3}\Big)~.\label{Lagm}
\end{align}
\( S_G \) is the action for the gravitational section, and \(S_M\) is the action for the matter fields. In the gravitational section, \(R\) is the Ricci scalar and $\Lambda=-3/L^2$ is the negative cosmological constant, with \(L\) the Ads radius. In the matter section, $F_{\mu\nu}=\nabla_{\mu}A_{\nu}-\nabla_{\nu}A_{\mu}$ is the strength of the U(1) gauge field. \(\rho_\mu\) is a complex vector field with mass \(m_p\) and charge \(q_p\), with its field strength defined as \(\rho_{\mu\nu} = D_{\mu}\rho_{\nu}-D_{\nu}\rho_{\mu}\), where the covariant derivative is given by $D_{\mu}\rho_{\nu}=\nabla_{\mu}\rho_{\nu}-i q_p A_\mu\rho_{\nu}$. The last two potential terms with coefficients \(\lambda\) and \(\tau\) introduce 4th and 6th power nonlinear self-interactions for the bulk vector field \(\rho_\mu\), which play a crucial role in realizing the various phase transitions beyond the second order.

In order to focus on the 4th and 6th power terms, we take the probe limit in the rest of this paper. Without the back reaction from the matter sector, the background metric takes the analytical form of (3+1)-dimensional Schwarzschild Ads black brane given as follows:
\begin{align}
&\ ds^{2}=\dfrac{L^2}{z^2}\left(-f(z)dt^{2}+\frac{1}{f(z)}dz^{2}+dx^{2}+dy^{2}\right)~,\label{metric1}
\end{align}
where the function \( f(z)\) is expressed as:
\begin{align}
&\ f(z)=1-\left(z/z_h\right)^3~.
\end{align}
In this metric, z represents the radial coordinate in the bulk, where $z=0$ corresponds to the Ads boundary
and $z=z_{h}$ denotes the horizon.
The Hawking temperature of this black brane solution is given by
\begin{align}
T= \frac{3}{4\pi z_h}~.
\end{align}

We set the ansatz for the matter fields as
\begin{align}\label{ansatz}
\rho_{x}=\psi(z)~, \quad A_{\mu}dx^{\mu}=A_{t}(z)dt=\phi(z)dt~,
\end{align}
and deduce the following equations of motion
\begin{align}
\psi''+\dfrac{f'}{f}\psi'-\dfrac{m_p^{2}}{z^{2}f}\psi+\dfrac{q_p^{2}\phi^{2}}{f^{2}}\psi&\nonumber\\
-\dfrac{2L^2\lambda_p}{f} \psi^{3}-\dfrac{3\tau z^{2}L^4}{f}\psi^{5}&=0,\label{eqpsi}\\
\phi''-\dfrac{2L^2q_p^{2}\psi^{2}}{f}\phi&=0~.\label{eqphi}
\end{align}
In order to solve the two coupled 2nd order differential equations, 4 sets of boundary conditions should be imposed either at the horizon or at the AdS boundary. The asymptotic behavior of the two bulk fields near the horizon \(z\to z_h\)  is given by the following expansions:
\begin{align}
    &\phi(z)=\phi_{0}+\phi_{1}(z-z_{h})+\mathcal{O}((z-z_{h})^{2})~,\\
    &\psi(z)=\psi_{0}+\psi_{1}(z-z_{h})+\mathcal{O}(z-z_{h})~.
\end{align}
Similarly, near the Ads boundary \(z\to0\), the expansions for the two bulk fields are
\begin{align}
	&\phi(z)=\mu-z \rho+...~,\qquad \\
	 &\psi(z)=z^{\Delta_{p-}}\psi_{p-}+{z}^{\Delta_{p+}} \psi_{p+}+...~,
\end{align}
where \(\mu\) and \(\rho\) represent the chemical potential and charge density in the dual field theory. For simplicity, we employ the standard quantization scheme, which means that \(\psi_{p-}\) corresponds to the source and \(\psi_{p+}\) corresponds to the expectation value of the vector operator. $\Delta_{P\pm}=\Big(1\pm\sqrt{1+4m_p^2}\Big)/2$ denote the conformal dimensions of \(\psi_{p\pm}\).  

We impose the 4 sets of boundary conditions as follows. The first is $\phi_{0}=0$ which ensures the finite norm of the gauge field $A_\mu$ on the horizon. The second is a natural linear relation between $\psi_{0}$ and $\psi_{1}$ for the finite branch of solution on the horizon. The third one is the source-free boundary condition \(\psi_{p-}=0\). Finally, we take a fixed value of the charge density $\rho$ to get the specific solution in the canonical ensemble.

With the above 4 sets of boundary conditions, there always exists a trivial solution with $\psi(z)=0$ and $\phi(z)=\rho(1-z)$, which is dual to the normal state without spontaneous breaking of the U(1) symmetry. In addition, when the charge density $\rho$ is larger than a critical value $\rho_c$, the single branch of solutions bifurcates to form a new branch with the spontaneous breaking of the U(1) gauge symmetry, where the vector expectation value grows from zero at the critical point. At higher eigenvalues $\rho_{ci}$, new branches of excited solutions with one or more nodes in $\psi(z)$ also grow up. Usually, the exited states are unstable and we only need to focus on the most stable solution between the normal solution and the condensed solution without node, which is confirmed by the thermodynamic potential. We take the fixed value of the charge density $\rho$, which means that we work in the grand canonical ensemble. Therefore, we derive the free energy by evaluating the renormalized Euclidean on-shell action. Within the probe approximation, the matter contribution is infinitesimal with respect to the gravity part. However, the metric in the probe limit is always the same for various solutions and the only difference in the free energy comes from the matter part which is given by
\begin{align}
F=\frac{V_2 L^{2}}{T}(\frac{\mu\rho}{2}+\int_{0}^{z_h}(\frac{qp^2 \phi^2 \psi^2}{f}-\lambda_p\psi^4-2z^2\tau\psi^6)dz)~,
\end{align}
where \(V_2=\int dxdy\) refers to the volume of the boundary spatial manifold.

There are three sets of scaling symmetries in the equations of motion~(\ref{eqpsi},\ref{eqphi}), which is useful to rescale the parameters $z_h$, $L$ and $q_p$ to any finite value.
\begin{flalign}
  (1).~& \phi(z)\rightarrow {\lambda}\phi(z),\psi(z)\rightarrow {\lambda}\psi(z),q_{p}\rightarrow {\lambda^{-1}}q_{p},&\nonumber\\&
    {\lambda_p}\rightarrow {\lambda^{-2}}{\lambda_p},{\tau}\rightarrow {\lambda^{-4}}{\tau}; & \\
  (2).~& \phi(z)\rightarrow {\lambda^{-2}}\phi(z),\psi(z)\rightarrow {\lambda^{-2}}\psi(z),f(z)\rightarrow {\lambda^{-2}}f(z),&\nonumber\\&L\rightarrow {\lambda}L,m_{p}\rightarrow {\lambda^{-2}}m_{p},{\tau}\rightarrow {\lambda^{2}}{\tau}; & \\
  (3).~& \phi(z)\rightarrow {\lambda^{-1}}\phi(z),\psi(z)\rightarrow {\lambda^{-1}}\psi(z),f(z)\rightarrow {\lambda^{-2}}f(z),&\nonumber\\
  & z\rightarrow {\lambda^{-1}}z,z_h\rightarrow {\lambda^{-1}}z_h. &
\end{flalign}
Therefore, in the rest of this work, we set \(z_h=L=q_p=1\) without loss of generality. We also fix the mass parameter \(m_p^2 =0\), and focus on the universal control from the 4th and 6th power terms with coefficients $\lambda$ and $\tau$.
\section{Various phase transitions from the universal control of the 4th and 6th power terms}\label{secPhaseTransition}
It is observed that in the holographic s-wave superfluid model, due to the precise control of the 4th and 6th power terms on the condensate curve, various phase transitions are realized in the probe limit~\cite{Zhao:2022jvs}. Further extension to include the back-reaction on the metric also confirm the same qualitative control from the 4th and 6th power terms~\cite{Zhao:2025tqq}. This feature is explained well from the landscape analysis, therefore it is expected that the control from the 4th and 6th power terms should be universal and also admit the various phase transitions in the holographic p-wave model in the probe limit. We illustrate typical phase transitions and give the quantitative parameter space which is useful for future studies based on the realization of the various phase transitions in an anisotropic superfluid in the probe limit.

Because we fix other parameters and only change the value of $\lambda$ and $\tau$ for the two nonlinear terms, it is obvious from the equations of motion~(\ref{eqpsi},\ref{eqphi}) that the (quasi) critical point always stays at the same point for the various phase transitions. This is because of that the condensate is infinitesimal right at the (quasi) critical point, therefore the nonlinear terms are always negligible. This is also an important aspect of the universal control from the two nonlinear terms that also holds in the holographic s-wave superfluids~\cite{Zhao:2022jvs,Zhao:2025tqq}.
\begin{figure}[h]
\centering
\includegraphics[width=0.45\columnwidth]{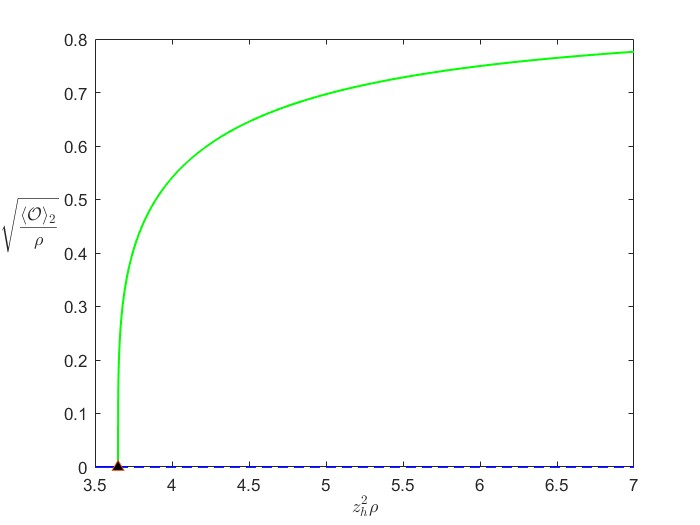}
\includegraphics[width=0.45\columnwidth]{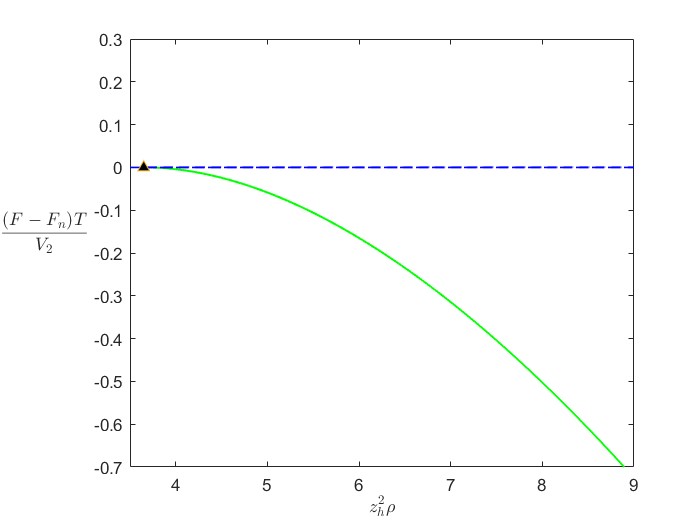}
\caption{The condensate (Left) and free energy (Right) of the second order phase transition with (\(\lambda=\tau=0\)).
The blue solid line represents the (meta-)stable section of the normal solution while the blue dashed line represents the unstable section of the normal phase beyond the (quasi) critical point \(\rho_c\). The solid green line represents the (meta-)stable p-wave superfluid phase. The black triangle indicates the location of the critical point with \(\rho_c\) = 3.6496, where the branch of the superfluid phase is connected to the branch of the normal phase.}\label{2nd}
\end{figure}
\subsection{Second-order phase transition}\label{SubSect2ndOrder}
In the probe limit, only the 2nd order phase transition is realized in previous studies on the holographic p-wave superfluid model, which is included in our setup as the special case of \(\lambda=\tau=0\). We represent the condensate curve as well as the free energy curve of this well studied case in the canonical ensemble with a fixed value of temperature $T$ in Figure ~\ref{2nd}. In these two plots, we can see a critical point at \(\rho_c= 3.6496\), below which the p-wave condensate is always zero. When the charge density is increased across the critical point, the system undergoes a second order phase transition, and the new phase with nonzero condensate gets the lowest value of free energy.

\subsection{Zeroth-order phase transition }\label{SubSect0thOrder}
An analytical study on the near critical region~\cite{Herzog:2010vz} show that a sufficiently negative value of the 4th power term in the holographic s-wave superfluid model will change the growing direction of the condensate curve, which results in a 1st-order phase transition or the global runaway instability~\cite{Zhao:2022jvs}. Ref.~\cite{Zhang:2021vwp} further elaborates on the influence of the quartic interaction term on the entire condensate curves in an s+p model, illustrating that nonzero \(\lambda\) significantly modifies the condensate behavior. Universal control of the 4th power terms is further confirmed in the holographic s+d model~\cite{Liao:2025knf}. 
\begin{figure}[b]
\centering
\includegraphics[width=0.45\columnwidth]{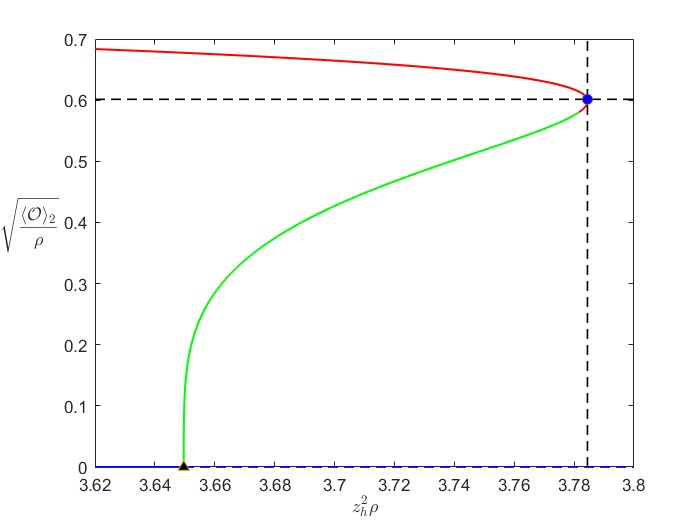}
\includegraphics[width=0.45\columnwidth]{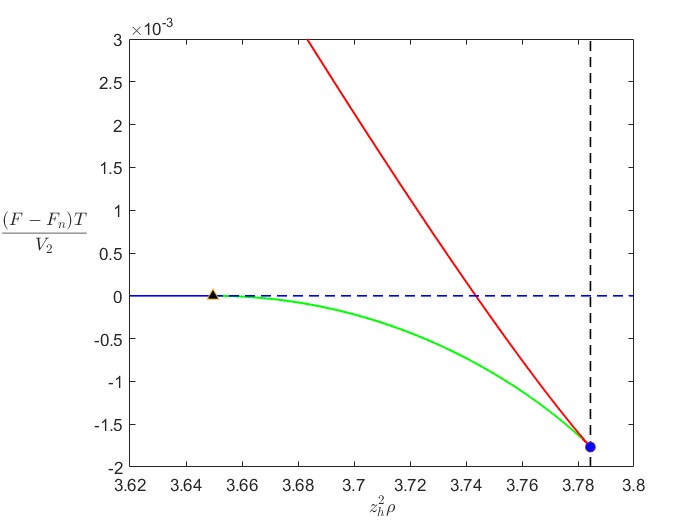}
\caption {The condensate (Left) and free energy (Right) of the zeroth- order phase transition with (\(\lambda=-0.7,\tau=0\)).
The notion of the green and blue lines as well as the black triangle are used as the same as that int Figure~\ref{2nd}. The solid red line represents the unstable section of the p-wave superfluid solutions, which is connected to the (meta-)stable section at turning point marked by the blue dot with \(\rho_t=3.78444\).}\label{0th}
\end{figure}

We confirmed that the universal control of the 4th power term is also valid in this p-wave model, which indicates that with a small negative value of $\lambda$, the p-wave condensate will turns back to the left at a finite condensate and results in the 0th order phase transition as shown in Figure ~\ref{0th} with $\lambda=-0.7$. The universal control also indicates that with a sufficiently negative value of $\lambda$, lower than $\lambda_s=-1.26$, the p-wave condensate will grows to the left on the critical point, just the same as its s-wave and d-wave cousins~\cite{Herzog:2010vz,Zhao:2022jvs,Liao:2025knf}.

\subsection{First-order phase transition}\label{subsection5}
In order to realize the first order phase transitions, we need the condensate curve to grow to the left at the critical point, which is available with a sufficiently negative value of $\lambda$ ($\lambda<\lambda_s=-1.26$). However, with only the negative 4th power potential term, the system suffers from the global instability from the landscape point of view. A simple solution is the including of higher power potential terms with positive coefficients, which will overcome the negative value of the fourth power term at very large condensate and preserve the free energy landscape bounded from below. For the simplest setup, we choose to add a 6th power term \(\tau|\psi|^6\) with coefficient $\tau$, which is proved to be successful in the holographic s-wave superfluid model~\cite{Zhao:2022jvs}. The quasi-normal modes as well as the dynamical evolution in such a setup are highly consistent~\cite{Zhao:2022jvs,Zhao:2023ffs}, providing a useful setup for investigating the 1st order superfluid phase transitions.

\begin{figure}[h]
\centering
\includegraphics[width=0.45\columnwidth]{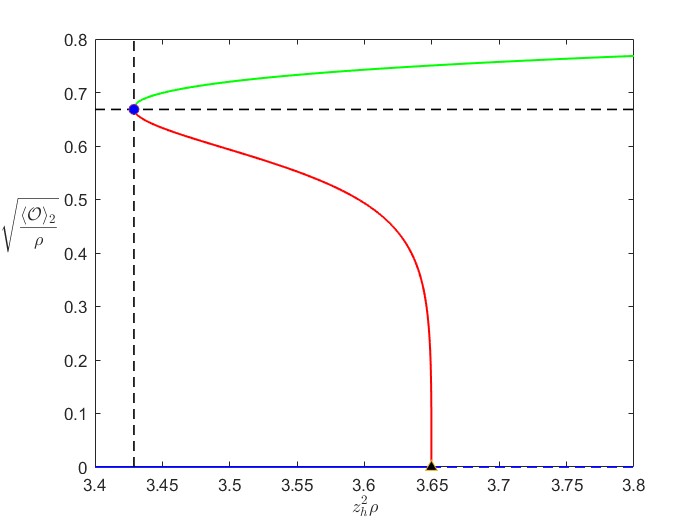}
\includegraphics[width=0.45\columnwidth]{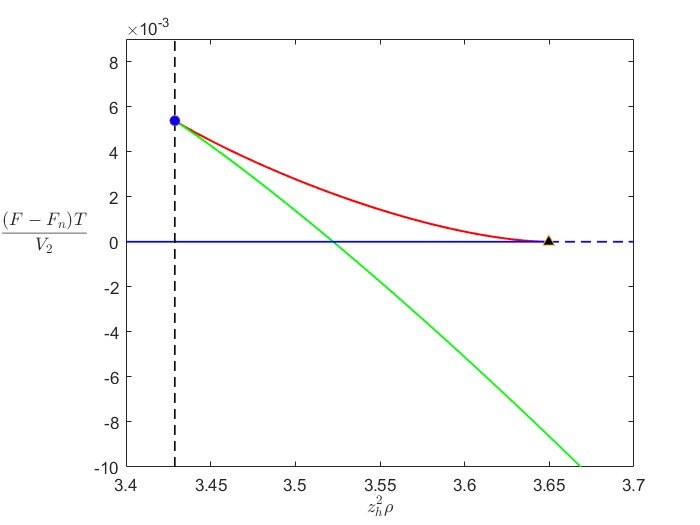}
\caption{The condensate (Left) and free energy (Right) of the first- order phase transition with (\(\lambda=-1.4,\tau=0.8\)).
The notion of the green and blue lines as well as the black triangle are used as the same as that int Figure~\ref{2nd}. In this plot, the unstable section of the p-wave superfluid solutions marked by the solid red line is connected to the normal solution on the quasi critical point marked by the black triangle, and is connected to the (meta-)stable section marked by the solid green line on the turning point marked by the blue dot with \(\rho_t=3.42887\).
}\label{1st}
\end{figure}

With the sufficiently negative value of $\lambda$ and a possitive value of $\tau$, the 1st order p-wave superfluid phase transitions are realized. We chosse $\lambda=-1.4$ and $\tau=0.8$ as a concrete example and plot the condensate as well as the free energy curves in Figure.~\ref{1st}. We can see that the condensate finally turns back to growing rightward, and the free energy curve show a typical swallow tail shape.
\subsection{Cave of wind, the 1st order phase transition into the superfluid solution with large condensate}\label{subsection6}
\begin{figure}
\centering
\includegraphics[width=0.45\columnwidth]{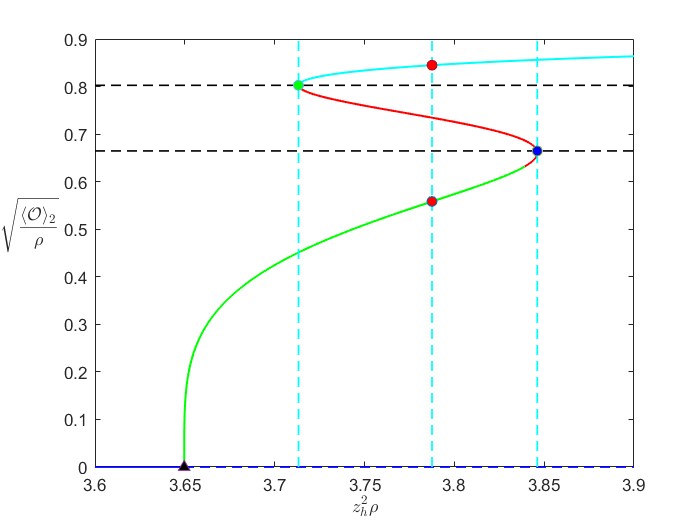}
\includegraphics[width=0.45\columnwidth]{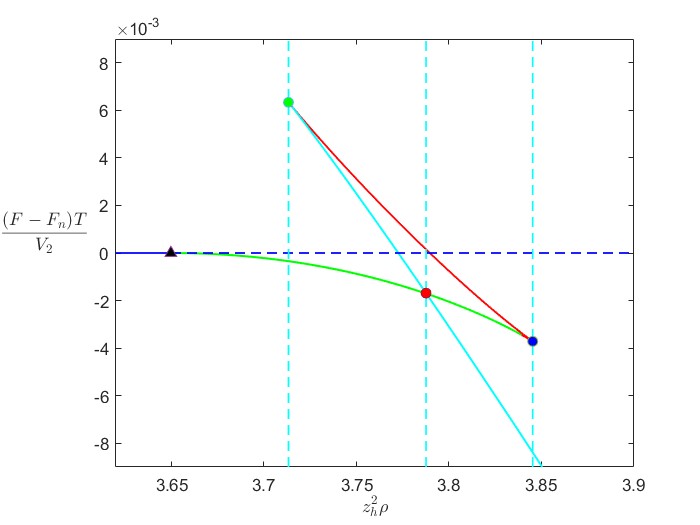}
\caption{The condensate (Left) and free energy (Right) of the COW phase transition with \(\lambda=-0.7\) and \(\tau=0.14\).
The notion of the green and blue lines as well as the black triangle are used as the same as that int Figure~\ref{2nd}. The (meta-)stable section of the p-wave superfluid solutions marked by the solid green line is connected to the normal solution on the quasi critical point marked by the black triangle, and is connected to the unstable section marked by the solid red line on the 1st turning point marked by the blue dot with \(\rho_{t1}=3.84522\). The other (meta-)stable section of the p-wave superfluid solutions marked by the solid cyan line is connected to the other side of the solid red line on the 2nd turning point marked by the green dot with \(\rho_{t2}=3.71325\). The red dot with \(\rho_T=3.78758\) mark the position of the first order phase transition between the two sections of superfluid solutions, which is identified by the intersection point of the free energy curve for the two sections of superfluid solutions.}\label{cow}
\end{figure}

The name ``cave of wind'' comes from the twisted pattern of the condensate curve, which is in principle a first order phase transition from the normal phase (when $\tau$ is very small) or the superfluid phase with lower condensate (when $\tau$ is relatively larger) into the superfluid phase with larger condensate. With only the 4th and 6th power terms, the ``cave of wind'' is realized with a small negative value of $\lambda$ ($\lambda_s<\lambda<0$) and a small positive value of $\tau$. We take a relatively larger value of $\tau$, when the charge density is increased from below the critical point $\rho<\rho_c$, the system first undergoes a second order superfluid phase from the normal phase, and later undergoes a first order phase transition from the superfluid phase with lower condensate to the superfluid phase with higher condensate. We illustrate the condensate as well as the free energy curves of such an example with $\lambda=-0.7$ and $\tau=0.14$ in Figure~\ref{cow}.

In Figure~\ref{cow}, we can see that as the condensate grows from the critical point, it first grows rightward and reached the first turning point at $\rho_{t1}=3.84522$ and then grows leftward. Soon it reached the second turning point at $\rho_{t2}=3.71325$ and finally grows rightward again. The free energy curve show that the first order phase transition is expected to take place on the phase transition point $\rho=\rho_T=3.78758$, and a typical swallowtail shape is presented.

\subsection{The \(\tau-\lambda\) parameter space for the various phase transitions} \label{subsection1}
\begin{figure}
\centering
\includegraphics[width=0.9\columnwidth]{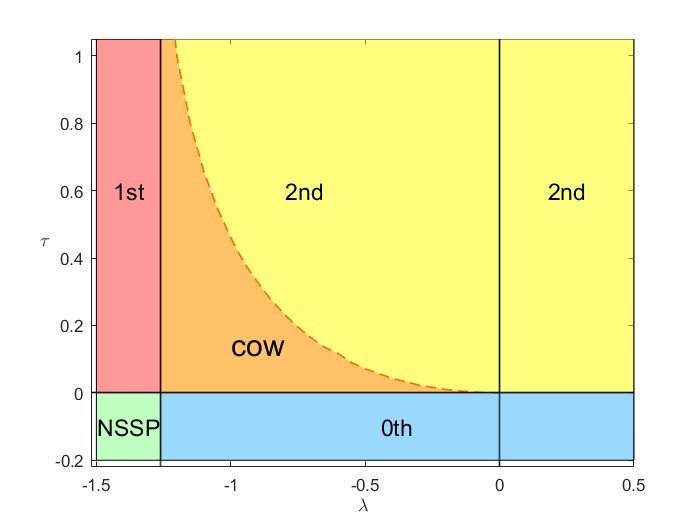}
\caption{The $\tau-\rho$ parameter space for the various phase transitions. In this plot, the symbol ``COW'' stands for "cave of wind" phase transition and the symbol ``NSSP'' stands for "no stable superfluid phase". ``2nd'', ``1st'' and ``0th'' denote the parameter space of the first-order phase transitions, the second-order phase transitions, and the zeroth-order phase transitions, respectively.}\label{lambda-tau}
\end{figure}

The above analysis on the various phase transitions confirmed the universal control from the 4th and 6th power terms on the p-wave superfluid phase transitions. In a word, the value of lambda controls the growing direction of the condensate curve near the critical point, while a positive value of tau ensures the final rightward growing of the condensate curve as well as the global thermodynamic stability. Now we give the quantitative parameter space including the various phase transitions in Figure~\ref{lambda-tau}.

We can see from Figure~\ref{lambda-tau} that below the special value $\lambda_s=-1.26$, the phase transition is first order with a positive value of $\tau$, while with a negative value of $\tau$, there is no stable superfulid phase (NSSP) because that the condensate curve always grows leftwards. When $-1.26=\lambda_s<\lambda<0$, the condensate curve grows rightward on the critical point, ensuring a second order phase transition from the normal phase. When both $\lambda$ and $\tau$ are positive, the condensate curve will never turn back, and the previous 2nd order phase transition is the only phase transition in this case. When $\lambda$ get a small negative value in the region $\lambda_s<\lambda<0$, with a sufficiently large value of $\tau$, larger than the value indicated by the dashed curve between the yellow and orange region, the condensate curve will also grow rightward forever, and there is also only the 2nd order phase transition. The condensate curve will turn back to be leftward and never turned to rightward again when $\tau$ is negative. Although this case indicates the global instability, we mark it as the ``0th order'' phase transition as in the previous studies, where the section of superfluid solutions with lower condensate is meta-stable. In the case marked by the orange region, $\lambda$ get a small negative value in the region $\lambda_s<\lambda<0$ while $\tau$ is positive but small. The small positvie $\tau$ ensure the final rightward growing of the condensate curve, but is not large enough to eliminate the leftward section at small condensate value, therefore the system show the ``COW'' condensate curve, indicating a 1st order phase transition into the section of superfluid solutions with large condensate.

This plot of the parameter space clearly represents the universal control of the 4th and 6th power terms and is very useful for future studies based on the realization of the various phase transitions.
\section{The $\tau-\rho$ parameter space for the various phase transitions}\label{secParameterSpace}
In the previous section, we confirmed the universal control from the 4th and 6th power terms and give the detailed parameter space for the various phase transitions. These results are very useful for future studies base on the realization of the various phase transitions. As a concrete example, we present a $\tau-\rho$ phase diagram displaying the supercritical region of the superfluid phase where the lower condensate section and higher condensate section of the superfluid phases are no longer distinguishable. The key point to realize such a phase diagram is to fix the value of $\lambda$ to an appropriate value in the region $\lambda_s<\lambda<0$, such as $\lambda=-0.05$.

\begin{figure}[b]
\centering
\includegraphics[width=0.9\columnwidth]{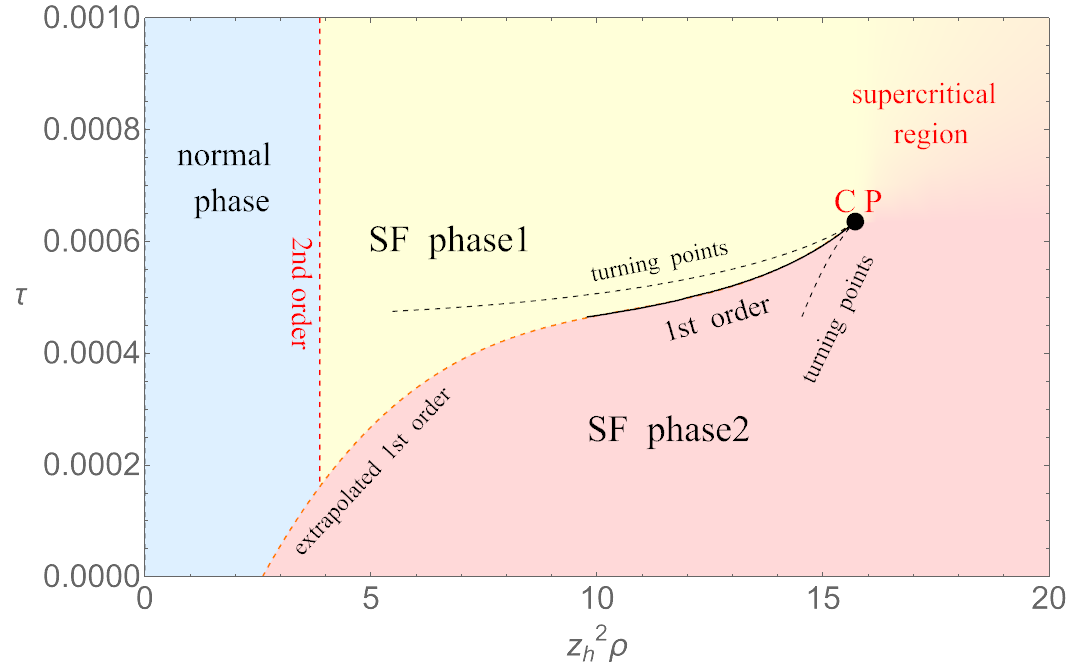}
\caption{The \(\tau-\rho\) phase diagram with \(\lambda = -0.05\). In this plot, the red dashed line delineates the second-order superfluid phase transition from the normal phase (blue region) to the section of the superfluid phase with lower condensate marked by the light yellow region. The black dot represent the critical point(CP) at the end of the black solid line indicating the 1st order phase transition points between the section of the superfluid solutions with lower condensate marked by the light yellow region and the section with larger condensate marked by the light red region. 
The dashed orange dashed line is interpolated from the black solid line because that the numerical calculation becomes difficult to get the the superfluid solutions with very large condensate. The two black dashed lines on the right and left sides of the solid black line represent the location of the 1st and 2nd turning points in the case of the COW phase transitions, respectively.}\label{rho-tau}
\end{figure}

The phase diagram with $\lambda=-0.05$ is plotted in Figure~\ref{rho-tau}. In this phase diagram, we can see a critical point (CP) with $\rho_{cp} = 16.1891$ and $\tau_{cp} = 0.000641$ at the end of the black solid curve which is made up of the first order phase transition points between the section of superfluid solutions with lower condensate and the section of superfluid solutions with larger condensate. Beyond the critical point is the supercritical region, where the two sections of superfluid solutions are no longer distinguishable. The two dashed black lines on the left and right sides of the solid black line indicate the turning points of the condensate curves, which are also the tips of the swallowtails on the free energy curves. Due to limitations of the numerical calculation, the lower section of the black solid curve is extrapolated to give an intuitive insight into the structure of the complete phase diagram. Recently, possible new boundaries in the supercritical region have become an interesting topic in the studies of black hole physics~\cite{Zhao:2024jhs,Zhao:2025ecg,Xu:2025jrk,Li:2025tdd,Wang:2025ctk,Li:2025lrq}. The easily realized phase diagram including the supercritical region is a good choice for further studies on the anisotropic supercritical superfluids.
\section{Conclusions and outlooks}\label{secConclusion}
In this study, we setup a holographic p-wave superfluid model with 4th and 6th power nonliear terms. We realized the varous phase transitions including the 2nd order, 1st order and 0th order ones and confirmed that the qualitative control of the 4th and 6th power terms on the superfluid phase transitions share the qualitative universality with the results discovered in the holographic s-wave model. We further present a $\tau-\rho$ phase diagram including the supercritical region to illustrate the powerful control from the 4th and 6th power terms, which useful in future studies involving the supercritical regions~\cite{Zhao:2024jhs,Zhao:2025ecg}.

Our research also provides a systematic way to realize various phase transitions in the holographic p-wave superfluids. Based on these phase transitions and the precise control from the 4th and 6th power terms, it is convenient to further extend the studies on the dynamical stability, dynamical evolution and inhomogenous configurations in the first order phase transitions~\cite{Zhao:2022jvs,Zhao:2023ffs}, to the case of p-wave superfluids. Including the back reaction on the metric based on this holographic model presenting rich phase structure is also a wise choice to further investigate important topics such as the holographic entanglement entropy as well as the black hole interior.

We should also notice that the two nonlinear terms preserve the symmetry between the p-wave solution and the p+ip solution in the probe limit~\cite{Nie:2016pjt}. Therefore, it is also interesting to combine these two nonlinear terms with various gravity background~\cite{Nie:2020lop} as well as magnetic effects~\cite{Yang:2021mit} to explore interesting features in the competition between the two configurations of vector condensate.
\section*{Acknowledgements}
YPW would like to thank Xing-Kun Zhang, Xin Zhao and Ru-Qing Chen for helpful discussions. This work was supported by the National Natural Science Foundation of China (Grant Nos. 12575054 and 11965013). ZYN is partially supported by Yunnan High-level Talent Training Support Plan Young \& Elite Talents Project (Grant No. YNWR-QNBJ-2018-181).
\bibliographystyle{apsrev4-1}
\bibliography{Reference}
\end{document}